# Mechanics of collapsing cavitation bubbles

Leen van Wijngaarden

University of Twente, Physics of Fluids group,and J.M.Burgers Centre for Fluid Dynamics,

P.O.Box 217, 7500AE Enschede, The Netherlands

**Abstract.** A brief survey is given of the dynamical phenomena accompanying the collapse of cavitation bubbles. The discussion includes shock waves, microjets and the various ways in which collapsing bubbles produce damage.

## 1.Bubble growth and collapse

In this special issue "Cleaning with bubbles "I will discuss the mechanical aspects of gas bubble implosions in liquids. They are essential for the subject" cleaning with bubbles " but have been studied in the past mainly in connection with cavitation in the flow along the blades of ship propellers and with damage caused by their implosion. A bubble can be produced in many ways: by vaporous growth due to low ambient pressure, as in propeller cavitation, by low pressures due to viscous stresses as occurs in the synovial liquid in your knees when you rapidly stretch these, by introducing locally high energy in the liquid, for example with a laser pulse, or by strong acoustical beams as in machines that pulverize kidney stones. In all these cases bubbles grow first, usually from microscopic size, and finally collapse due to the return of high pressure in their vicinity. The cleaning effect is coming from the final stage of the collapse and therefore it is worthwhile to go into this somewhat further. The earliest studies were undoubtedly in the field of cavitation on ship propellers and we will deal with that example. Rayleigh`s [1] mathematical formulation of the collapse of a cavity is

$$R\ddot{R}+\frac{3}{2}(\dot{R})^2 = \frac{p(R)-p(\infty)}{\rho} \qquad (1)$$

In this equation $R$ is the radius of the bubble, $\rho$ the density of the liquid and $p$ the liquid pressure, $p(R)$ at the interface and $p(\infty)$ far away from the bubble. A dot means differentiation with respect to time. With initial radius $R_0$ this can, for an empty cavity, $p(R)=0$, be integrated to give

$$\dot{R}^2 = \frac{2p(\infty)}{3\rho}\left(\frac{R_0^3}{R^3}-1\right) \qquad (2)$$

In the final stage of the collapse $R_0$ is much larger than $R$ and the bubble wall velocity $u$ reaches the value

$$u = \left\{\frac{2p(\infty)}{3\rho}\left(\frac{R_0}{R}\right)^3\right\}^{\frac{1}{2}} \qquad (3)$$

Liquid rushes into the cavity and with an empty cavity there is a singularity at the end. In reality the bubble contains gas, air mainly, and vapor. These are compressed, condensation may take place at the same time, and the collapse stops finally, at a radius of the bubble so small, that pressures of thousands of atmospheres could be reached provided the bubble stays spherical. To give an idea, suppose a bubble starts to collapse under a pressure difference of one bar far away and a pressure of *0,01* bar internal gas pressure at an initial radius of *1 mm*. With a polytropic constant (see below) of *1.4,* a minimum radius is reached of *15 μ* and the final pressure is about *3 .10$^7$N/m$^2$*. The time that it takes to reach the minimum radius is of order of *10$^{-4}$sec,* using (3) as an estimate. Since Rayleigh [1] many investigations have been conducted to take into account the various physical phenomena that affect the growth and collapse of spherical bubbles. Good accounts are given in the review papers by Plesset and Prosperetti [2], and by Blake and Gibson [3]. There are in addition several books describing the fundamentals of bubble cavitation, such as Young[4], Brennen[5] .

Most developments took place in the second half of the twentieth century. Instead of (1) is nowadays used an equation called the Rayleigh-Plesset equation, which takes account of surface tension , viscous stresses, gas and vapor content. So, the pressure *p (R )* is the pressure due to the gas content of the bubble, which under compression may behave from isothermal to adiabatic. In the first case the polytropic constant *κ* in the relation *p( R )/p$_0$ = $(\rho(R)/\rho_0)^\kappa$* is equal to one and in the second case *1,4* , for air. In the example above the value *4/3* is used. The Rayleigh –Plesset equation is

$$R\ddot{R}+\frac{3}{2}(\dot{R})^2 = \frac{1}{\rho}\left\{p(R) - p(\infty) - \frac{2\sigma}{R} - \frac{4\mu}{R}\dot{R}\right\} \quad (4)$$

In this equation two properties of the liquid occur apart from the density, the surface tension *σ* and the dynamic viscosity *μ*. The terms representing their effect on the collapse are *2σ/R* and *4μ/R $\dot{R}$* respectively. The value of *σ* for a water-air interface is *0,07 N/m* and is therefore only important for very small bubbles. The value of *μ* is *10$^{-3}$ kg/ms* . Viscous dissipation is part of the damping of bubble motions but is in general overruled by dissipation due to heat conduction from the bubbles into the surrounding fluid and acoustic radiation. Therefore the dynamic equation (4) is not sufficient to determine gas pressure and other properties of the collapse, but should be accompanied by an energy equation. In the early days of bubble research it was assumed that the gas inside bubbles behaves either isothermally or adiabatically, depending on the thickness of the thermal boundary layer inside the bubble as compared to the radius. This leads to only very crude approximations. To obtain reliable predictions, the full set of dynamic and energy equations should be applied. Analytic solution is then in general not possible, and numerical analysis is needed.  To clarify the complex connection between thermodynamic and dynamic behavior of bubble growth and collapse important contributions have been made by Prosperetti and co-workers see e.g. Watanabe & Prosperetti[6]. High gas pressures as mentioned as an numerical example in connection with equation (4) are not fully reached in practice , because collapsing bubbles do not remain spherical. The collapse is unstable, due to the curvature of the interface. Note that a flat interface is stable when the heavy liquid accelerates in to the lighter one. The bubble surface is deformed while the gas inside is compressed. In Figure1 is shown how strongly deformed the bubble surface becomes.

Figure 1

In the case of unconfined surroundings there is finally a rebound. Pressure waves are sent out from the bubble in to the ambient liquid. These are so strong that shock waves are formed in the liquid, as shown in Figure 2

Figure 2

In this stage of the collapse the compressibility of the liquid becomes important .In the dynamic equation (4) the liquid is supposed to be incompressible. Extension of the equation (4) to the compressible case then is needed and has been done by several authors, see Lezzi and Prosperetti [8].

**2. Bubble collapse close to a boundary.**

New phenomena occur when the collapse takes place near a boundary. We will concentrate on a boundary which may be considered as rigid. Whether a piece of textile , to be cleaned for example, may be considered as such, depends on the mechanical properties. Into a rigid boundary fluid cannot penetrate. When a bubble is collapsing close to a wall , fluid motion is induced in the wall with a component perpendicular to the wall. This must be annihilated because of the above mentioned boundary condition. This is achieved by a virtual image of the bubble in the wall. This image induces velocities in the fluid above the wall ,directed towards the center of the image bubble, see Figure 3.

Figure 3.

The fluid acquires momentum in the direction of the wall . The momentum of the fluid appears not to be a practical quantity. A useful quantity is the so-called impulse of the bubble ,which is representative of the inertia of the fluid in the vicinity of the bubble moving towards the wall. Mathematically the impulse $I$ is expressed as

$$I = UM, \qquad (5)$$

where $U$ is the velocity of the centre of the bubble in the direction normal to the wall, and $M$ is the so-called added mass of the bubble, a measure of the mass of liquid accelerated together with the bubble. The added mass is proportional to the density $\rho$ of the fluid and to the volume of the bubble, $M = C\rho V$ , where $C$ is a constant. Hence $I = C\rho V U$ . The advantage of the concept of impulse, associated with the bubble motion, is that it remains ,in the absence of external forces, approximately constant during the approach of the bubble towards the wall see for a detailed argumentation Benjamin and Ellis [11] . Since the volume of the collapsing bubble decreases , $U$ must increase. But the bubble becomes so small that this increasing velocity cannot be accommodated in the motion of a small sphere. Inside the sphere is a uniform gas pressure and this together with surface tension cannot support the large external pressure differences associated with high velocities. What happens is that the bubble surface suffers an involution whereby the upper surface folds into the interior of the bubble and a microjet is formed which pierces the side of the bubble which is closest to the wall. This is shown in Figure 4 which is taken from [9] .

Figure 4

The velocity of the liquid forming the microjet can reach such high values that severe damage is inflicted to the wall material. The jet formation was suggested as a source of damage by Kornfeld and Suvorov[10] and a pioneering paper on the role of the impulse is Benjamin and Ellis [11]

**3. What causes the damage to material , resp. the cleaning of textile.**

From the discussion in the foregoing sections two sources for destructive action of collapsing cavitation bubbles emerge, shockwaves emitted just after the first rebound of a collapsing bubble and microjets formed by bubbles collapsing in the vicinity of a sold wall. Although much research took place during the past 50, or so, years, it is not clear which one is dominant. Experiments as for example in [7] and [13] are not conclusive in this respect. We give here some arguments in favor and against both phenomena. High pressures due to shock waves are probably dominant in collapsing bubble clouds. Until now we considered single bubble collapse ,but in many circumstances , for example with acoustically generated bubbles , there is a collection of collapsing bubbles. In such a cloud the collapse of one bubble may reinforce that of neighboring ones. In that way high pressures are established over an area much larger than the size of an individual bubble. In clouds microjets are less effective. As explained in the previous section , they occur when the bubble has acquired an initial impulse. With a single bubble moving towards a wall the impulse is directed normal to the wall. In a cloud the attraction of a neighboring bubble may be much stronger than that by the wall, with the consequence that the microjet is directed not to the nearby wall but to a bubble in its vicinity. On the other hand with a single bubble the shock strength decreases rapidly with distance from the wall, so that only bubbles close to the wall are effective in producing damage. In that case microjets are probably more important. Till thus far the wall has been assumed rigid. In the case of textile this may not be the case .The opposite of a rigid wall is a completely soft wall, where the boundary behaves as a constant pressure surface . Then the image of a collapsing bubble is an expanding bubble , working as a source to annihilate the pressure due to the collapsing bubble, which is a sink. The impulse is directed away from the wall. In that case microjets are also directed away from the wall, and shock waves are more effective for cleaning.

**3. Conclusion.**

In the paper it is argued that in the process of "Cleaning with Bubbles" the cleaning effect is due to the phenomena shock wave generation and microjet formation , both occurring at the final stage of bubble collapse.

Captions of Figures 1-4

Figure 1

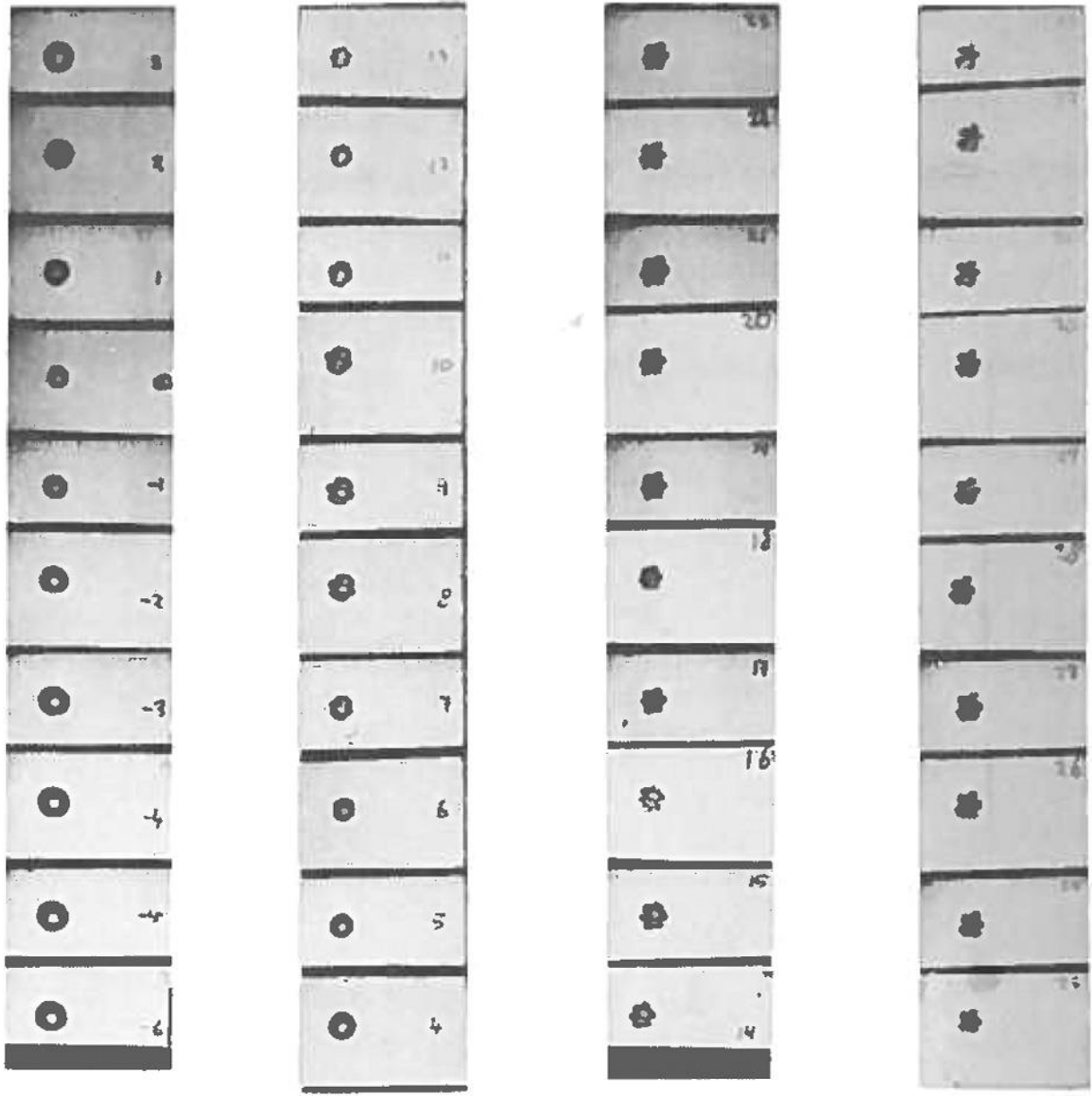

Collapsing bubble. The initial radius is *1.0 mm*. The pressure difference under which the bubble collapses is *0.75 bar*. The time difference between successive frames is *40μs*. From [12]

Figure 2

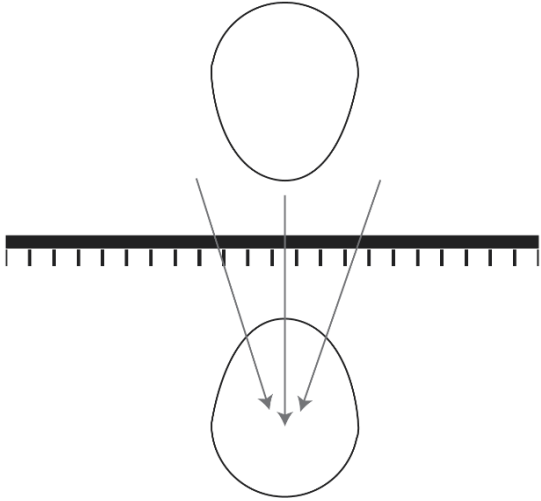

Shock waves emitted from bubbles. *a)* Bubbles far away, compared with their radius, from a boundary, *b)* close to the boundary ( note the beginning of a microjet in *b*). From [7]

Figure 3

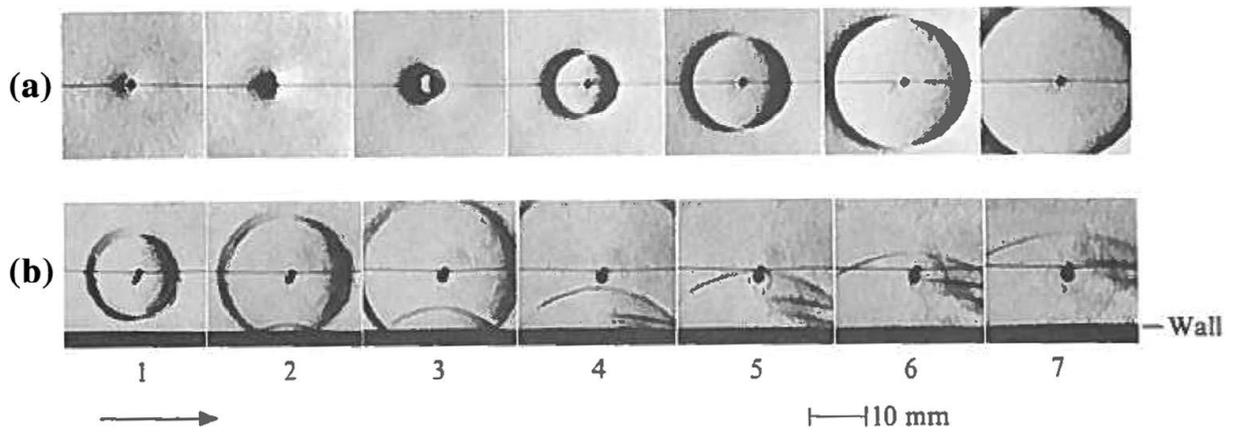

Bubble collapsing close to a rigid boundary

Figure 4

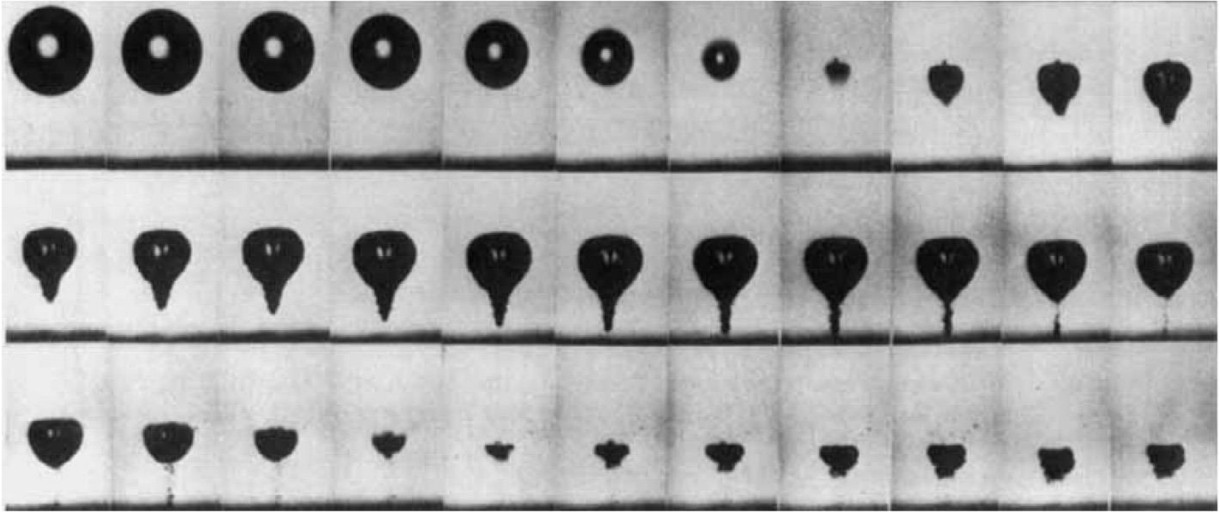

Formation of a microjet from a collapsing bubble close to a boundary. From [9]